\documentclass[11pt]{article}

\usepackage[final]{acl}

\usepackage{times}
\usepackage{latexsym}

\usepackage[T1]{fontenc}

\usepackage[utf8]{inputenc}
\usepackage[table,xcdraw]{xcolor}

\usepackage{microtype}

\usepackage{inconsolata}

\usepackage[table]{xcolor}
\usepackage{multirow}
\usepackage{titletoc}
\usepackage[most]{tcolorbox}

\usepackage{xcolor}
\usepackage{listings}
\usepackage[most]{tcolorbox}
\tcbuselibrary{breakable,skins,listings}

\newtcolorbox[auto counter, number within=section]{promptbox}[2][]{%
  colback=white,
  colframe=green!50!gray!40!black,
  width=\columnwidth,
  arc=1mm,
  boxrule=0.5mm,
  title={\normalsize#2},
  #1
}

\usepackage{algorithm}
\usepackage{algorithmic}
\usepackage{multirow}
\usepackage{booktabs}
\usepackage{amsmath}
\usepackage{amssymb}
\usepackage{graphicx}
\usepackage{bbding}

\usepackage{newfloat}
\usepackage{listings}
\usepackage{pifont}

\definecolor{darkorange}{rgb}{1.0, 0.55, 0.0}
\definecolor{mygreen}{RGB}{0,180,0}
\definecolor{myred}{RGB}{180,0,0}  

\newcommand{\icohalf}{\textcolor{darkorange}{\ding{51}\kern-0.65em\ding{55}}}

\title{
One-Stage Multi-Task Instruction-Guided 3D Spatial Audio Editing
}

\author{
\textbf{Ke Lei\textsuperscript{1,*}},
\textbf{Chenyuhao Wen\textsuperscript{1,*}},
\textbf{Yu Zhang\textsuperscript{2,*}},
\textbf{Wenxiang Guo\textsuperscript{1}},
\textbf{Changhao Pan\textsuperscript{1}},
\textbf{Sashuai Zhou\textsuperscript{1}},
\\
\textbf{Yongshi Li\textsuperscript{1}},
\textbf{Ruiqi Li\textsuperscript{2}},
\textbf{Ruofan Hu\textsuperscript{1}},
\textbf{Haorui Xu\textsuperscript{1}},
\textbf{Xiang Yin\textsuperscript{2}},
\textbf{Zhou Zhao\textsuperscript{1,$\dagger$}}
\\
\\
\textsuperscript{1}Zhejiang University,
\textsuperscript{2}ByteDance
\quad
\textsuperscript{*}Equal contribution
\quad
\textsuperscript{$\dagger$}Corresponding author
\\
\href{mailto:leike@zju.edu.cn}{leike@zju.edu.cn}
\quad
\href{mailto:zhaozhou@zju.edu.cn}{zhaozhou@zju.edu.cn}
}

\begin{document}

\maketitle

\begin{abstract}
Spatial audio editing modifies an existing soundfield according to a user's instruction while preserving the rest of the scene. Unlike conventional audio editing, it must reason jointly about audio events, spatial information, dynamic changes, and environmental information in first-order Ambisonic (FOA) waveforms. Existing language-guided editors mainly target conventional audio or rely on sequential operations, and therefore do not directly support one-stage editing for complex 3D spatial instructions. We present \textsc{SwanWeave}, the first one-stage multi-task framework for instruction-guided 3D FOA spatial audio editing. We build paired FOA supervision from open-source speech and sound-effect corpora using controllable room simulation, covering more than ten single-operation and compound tasks across the four editing axes. To handle this heterogeneous edit space, \textsc{SwanWeave} uses Spatial Edit Mixture-of-Experts (SE-MoE) with dual-level routing, selecting task-aware expert combinations for compound instructions and frame-level routed/null experts for local edit decisions. We further introduce Spatial Preference Optimization (SPO), a Direct Preference Optimization (DPO)-based alignment objective with edit-specific negative targets, and adopt staged training to improve natural-language grounding. Experiments show that \textsc{SwanWeave} achieves better editing quality than existing general audio editors and spatial audio baselines across all tasks. 
Spatial audio editing demos can be found at \href{https://swanaigc.github.io/#swanweave}{swanaigc/swanweave/}, code can be found at: \href{https://github.com/MM-Speech/SwanWeave}{github.com/MM-Speech/SwanWeave}.
\end{abstract}

\section{Introduction}

\begin{figure*}[t]
    \centering
    \includegraphics[width=0.8\textwidth]{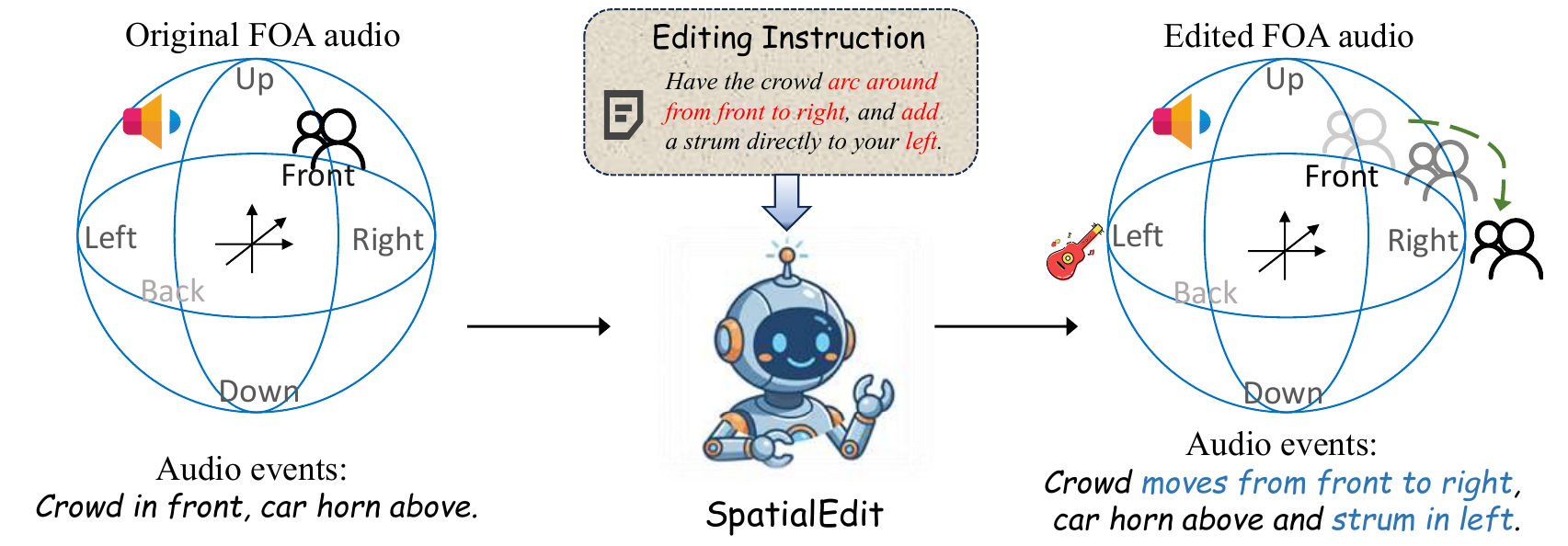}
    \caption{Overview of \textsc{SwanWeave}, which follows natural-language instructions to edit FOA audio, such as moving a crowd from the front to the right and adding a strum on the left while preserving unchanged events.}
    \label{fig:intro}
    \vspace{-0.5em}
\end{figure*}

Spatial audio is increasingly used in immersive media, including virtual and augmented reality, telepresence, games, and film production \citep{zhang2017surround,poeschl2013integration,broderick2018importance,2020-3-397}. It turns audio into a scene: sources occupy locations, move over time, and interact with the surrounding space. This makes editing especially useful. In production, a scene is often close to the desired result but still needs a targeted change to its content, source layout, motion, or room impression. Editing should therefore operate on an existing spatial scene, applying only the requested change instead of rebuilding the entire sound field.

Natural-language instructions provide a convenient interface for this setting, since users can describe the intended edit without specifying a full target scene.
Progress in audio-language models and spoken dialogue benchmarks has advanced audio-text representation learning and acoustic-language understanding \citep{hu2025vela,cheng2025voxdialogue}, and language-guided audio editing methods have made clear progress for conventional audio, where a model must ground the target event, modify it, and avoid drifting unedited regions \citep{wang2023audit,manor2024zeta,jia2024audioeditor}.
Spatial editing, however, is not just the same problem with more channels. As shown in Figure~\ref{fig:intro}, an instruction may target audio events, spatial information, dynamic changes, or environmental information, and these factors are coupled in the waveform. Moving a source farther away, for example, can change its energy, directionality, reverberation, and masking with other sources. Meanwhile, unedited sources should preserve their content, trajectories, and room-acoustic context. These requirements make instruction-guided spatial audio editing distinct from general audio editing.

Currently, spatial audio editing still faces two main challenges. 1) \textbf{One-stage generation.} Existing language-guided editing methods can synthesize an edited waveform, but they mainly target monaural or conventional audio operations and do not treat 3D spatial variables as editable controls \citep{wang2023audit,manor2024zeta,jia2024audioeditor}. SmartDJ is closer to spatial editing, but it operates on stereo audio and decomposes a high-level request into sequential atomic operations \citep{lan2025smartdj}. Such a pipeline does not directly learn the joint acoustic consequence of a compound spatial instruction, and errors may accumulate across stages. 2) \textbf{Task coverage.} Spatialization and spatial audio generation methods model directionality, localization, or video-conditioned spatial sound \citep{richard2021neural,leng2022binauralgrad,li2024tas,pan2025wild,liu2025omniaudio,kim2025visage}, and SALM provides representation-level spatial-language manipulation \citep{hu2025salm}. These methods are not waveform-level editors for existing first-order Ambisonic (FOA) scenes. To our knowledge, prior work has not conducted FOA editing covering difficult 3D and dynamic operations, including audio-event edits, 3D relocation, dynamic trajectories, and environmental changes.

To address these challenges, we present \textsc{SwanWeave}, the first one-stage multi-task framework for instruction-guided 3D FOA spatial audio editing. We build paired FOA supervision from open-source speech and sound-effect corpora using controllable room simulation, covering over ten single-operation and compound tasks across audio events, spatial information, dynamic changes, and environmental information. On the model side, we introduce Spatial Edit Mixture-of-Experts (SE-MoE), which uses dual-level routing to choose task-aware expert combinations for compound instructions and frame-level routed/null experts for local edit decisions, improving one-stage edit quality across heterogeneous tasks and events. We further introduce Spatial Preference Optimization (SPO), a Direct Preference Optimization (DPO)-based alignment objective with edit-specific negative targets, to improve instruction following for challenging spatial edits. Finally, we adopt a staged training strategy: text-to-FOA pretraining first builds semantic grounding for spatial audio, and single-operation and compound-edit training then gradually teach the model one-stage editing for complex compound tasks, thereby improving its understanding of natural-language instructions. Experimental results show that \textsc{SwanWeave} achieves better editing quality than existing general audio editors and spatial audio baselines across all tasks.

Our contributions are summarized as follows:
\begin{itemize}
    \item We introduce \textsc{SwanWeave}, the first one-stage multi-task framework for instruction-guided 3D FOA spatial audio editing, supported by paired FOA supervision over single-operation and compound tasks.
    \item We propose SE-MoE with dual-level routing, which combines instruction-level task experts with frame-level routed/null experts to improve one-stage edit quality across heterogeneous tasks and audio events.
    \item We introduce SPO, a DPO-based alignment objective with edit-specific negative targets, and a staged training strategy that improves natural-language grounding for complex spatial edits.
    \item Experimental results show that \textsc{SwanWeave} achieves better editing quality than existing baselines across all tasks.
\end{itemize}

\section{Related Work}

\subsection{Audio Editing}

Audio editing aims to modify an existing recording while preserving the parts that are not targeted by the edit. Recent text-to-audio models provide strong generative priors for open-domain sound synthesis \citep{kreuk2022audiogen,huang2023make,liu2023audioldm,liu2024audioldm,ren2025generative}, and language-guided editing methods build on such priors to generate an edited output from an input clip and a textual condition. AUDIT constructs triplets of input audio, instruction, and edited audio, and trains a latent diffusion model for adding, dropping, replacement, inpainting, and super-resolution \citep{wang2023audit}. ZETA and ZEUS explore zero-shot editing with DDPM inversion, using either text prompts or unsupervised semantic directions \citep{manor2024zeta}, while AudioEditor adapts diffusion-inversion techniques for training-free editing with attention to precise edits and preservation \citep{jia2024audioeditor}. SmartDJ is closest to our setting: it extends editing to stereo scenes and uses an audio-language-model planner to decompose high-level requests into sequential atomic operations executed by a diffusion editor \citep{lan2025smartdj}. These works establish important editing paradigms, but their edit spaces remain largely semantic or stereo-level. They do not jointly model 3D spatial attributes, dynamic trajectories, and room-acoustic transformations as editable targets. In contrast, \textsc{SwanWeave} treats audio events, spatial information, dynamic changes, and environmental information as a coupled edit space and directly generates the final edited 3D scene in one stage.

\subsection{Spatial Audio Generation}

Spatial audio research spans generation, spatialization, and vocoding, which synthesize or reconstruct audio with directional and environmental cues \citep{zhu2025asaudio,zhu2026csavocoder}. Early learning-based systems studied spatial sound from visual or geometric context, including visual sound localization and audio-visual navigation environments \citep{gao20192,chen2020soundspaces,chen2022soundspaces}. Binaural and text-guided spatialization methods convert mono or content-conditioned audio into spatial audio by modeling listener- or text-specified localization cues \citep{richard2021neural,leng2022binauralgrad,li2024tas,pan2025wild}; recent systems further generate spatial audio from multimodal inputs or in streaming settings \citep{liu2025omniaudio,kim2025visage,zhang2025isdrama,lei2026swansphere}. Another related line studies spatial audio representations and spatial audio-language alignment: spatial datasets and SELD-style annotations provide event and direction labels \citep{shimada2023starss23}, and SALM builds structured semantic and spatial embeddings that enable directional manipulation in representation space \citep{hu2025salm}. These methods are useful for spatial understanding, localization, spatialization, or generation, but they do not directly solve instruction-guided editing of an existing 3D sound scene. \textsc{SwanWeave} instead takes an existing spatial mixture, applies a requested audio-event, spatial, dynamic, or environmental edit, and preserves the rest of the scene.

\section{Method}

\subsection{Problem formulation}

Let $a_{\mathrm{src}}\in\mathbb{R}^{4\times T}$ and $a_{\mathrm{tgt}}\in\mathbb{R}^{4\times T}$ denote the source and target first-order Ambisonic (FOA) audio, respectively, where the four channels encode the spatial soundfield and $T$ is the waveform length. Given a natural-language editing instruction $c$, such as \emph{Insert a motorcycle sound behind on the left}, our goal is to learn an editor
\begin{equation}
    \hat{a}_{\mathrm{tgt}} = F_{\theta}(a_{\mathrm{src}}, c),
\end{equation}
so that the generated audio $\hat{a}_{\mathrm{tgt}}$ follows the requested edit while preserving unrelated sound events, spatial cues, and room characteristics.

Spatial editing can change what sounds, where it is located, how it evolves over time, or what acoustic space it occupies. We group the tasks along four axes:
\begin{itemize}
    \item \textbf{Audio events}: add, remove, replace, extract, enhance, or attenuate a sound event.
    \item \textbf{Spatial information}: edit the spatial attributes of a source, including azimuth, elevation, distance, and source layout.
    \item \textbf{Dynamic changes}: apply time-varying gain changes or dynamic spatial trajectories, such as angular motion and distance motion.
    \item \textbf{Environmental information}: change room-acoustic conditions such as perceived room size and reverberation while keeping the scene content and source layout consistent unless otherwise instructed.
\end{itemize}
These axes are coupled in the waveform. A compound instruction should therefore be learned as a direct source-to-target transformation rather than as a chain of separately executed operations. \textsc{SwanWeave} follows this formulation and directly predicts the final edited FOA scene in one stage.

\subsection{Data construction}
\label{sec:data_construction}

One-stage editing requires paired supervision in which the target already contains the complete acoustic consequence of the instruction. We construct such supervision from open-source speech and sound-effect corpora, including AudioCaps \citep{kim2019audiocaps}, FSDKaggle2019 \citep{fonseca2019audio}, PicoAudio \citep{xie2024picoaudio}, LibriSpeech \citep{panayotov2015librispeech}, and Spatial LibriSpeech \citep{sarabia2023spatial}. For each example, we sample two to four audio events to form a source scene and synthesize its target scene according to the requested edit. Controllable room simulation provides FOA rendering with specified directions, distances, motion patterns, and room-acoustic conditions \citep{scheibler2018pyroomacoustics}.

Each training item is a triplet $(a_{\mathrm{src}}, c, a_{\mathrm{tgt}})$ consisting of a source FOA waveform, an editing instruction, and the final target FOA waveform. The target is rendered after all requested changes have been applied, so the model can learn compound edits without observing intermediate operations. The task set covers more than ten single-operation and compound tasks across audio events, spatial information, dynamic changes, and environmental information. We generate about 125K triplets for each single-operation task and 250K triplets for compound editing; each example is approximately 8.9 seconds long. To reduce dependence on a fixed instruction style, we generate three paraphrases for each edit instruction using Gemini 2.5 Pro \citep{comanici2025gemini}. 
Additionally, in the test set, we include 50 examples for each single-operation edit type, and separately synthesize 50 examples each for two-operation, three-operation, and four-operation edits. The test set and training set do not share original audio clips to prevent data leakage.

The same data also supports the staged training curriculum described below. Before editing training, we decompose each source-target pair into scene-level FOA-caption examples, where the caption describes a complete spatial scene. These pairs initialize the model with text-to-FOA generation ability and with language grounding for events, spatial layout, motion, and room acoustics. Editing triplets are then used for single-operation training and, finally, compound-edit training.

\begin{figure*}[t]
    \centering
    \includegraphics[width=0.98\textwidth]{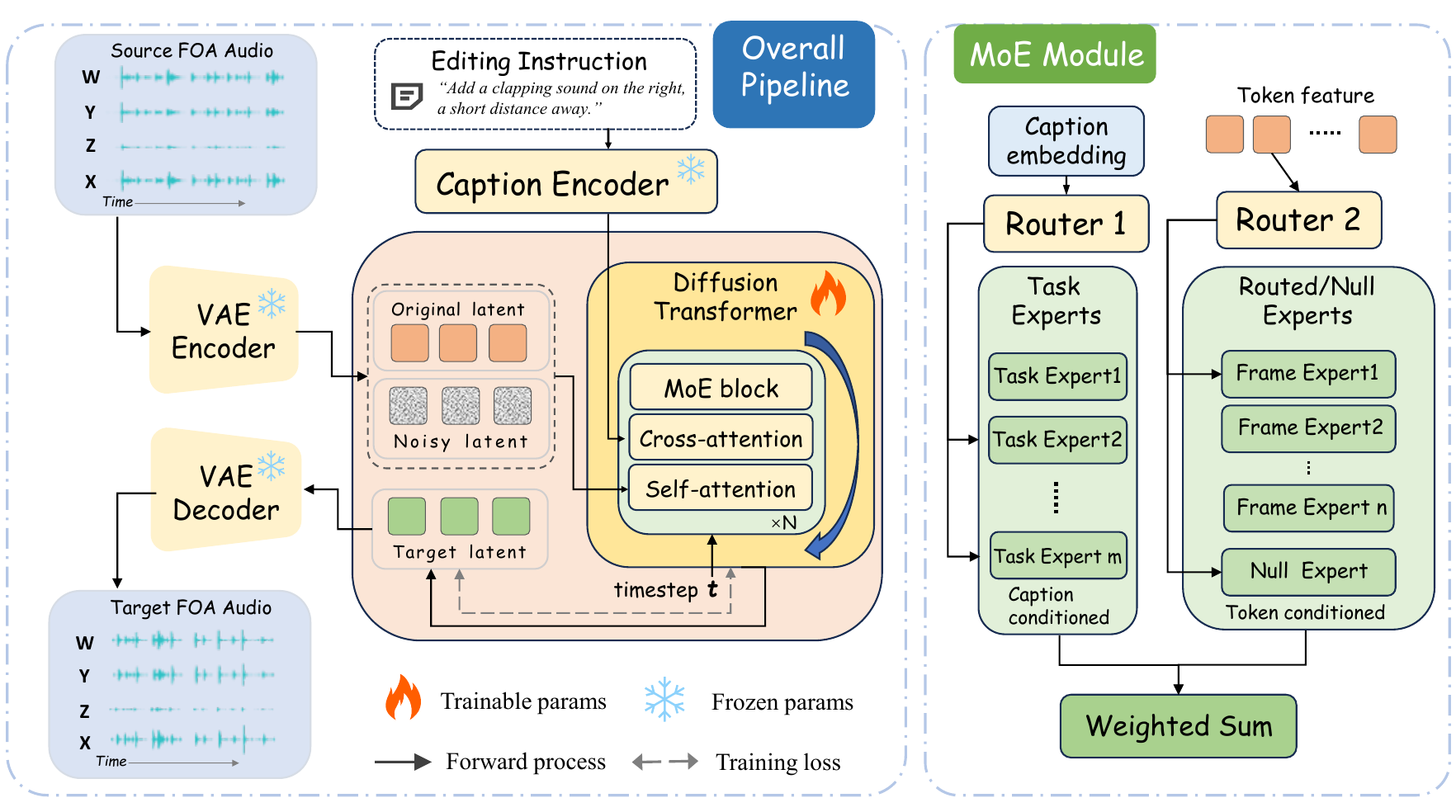}
    \caption{Overview of \textsc{SwanWeave}. The left panel shows the overall training and inference pipeline, and the right panel shows the SE-MoE module with instruction-level task routing and token-level routed/null experts.}
    \label{fig:wide}
    \vspace{-0.5em}
\end{figure*}

\subsection{Model overview}

Editing FOA waveforms directly would require the model to synthesize long four-channel signals while preserving fine spatial cues. We instead edit in a compact latent space, where the generator can focus on the requested transformation rather than low-level waveform reconstruction. As shown in Figure~\ref{fig:wide}, \textsc{SwanWeave} contains an FOA encoder, an instruction encoder, and a flow-matching latent editor equipped with Spatial Edit Mixture-of-Experts (SE-MoE). We use a stereo FOA VAE ~\cite{evans2025stable} to map spatial audio into a continuous latent space. The encoder maps source and target FOA waveforms into latent representations:
\begin{equation}
    z_{\mathrm{src}}=\mathcal{E}(a_{\mathrm{src}}), \quad
    z_{\mathrm{tgt}}=\mathcal{E}(a_{\mathrm{tgt}}),
\end{equation}
where $z_{\mathrm{src}},z_{\mathrm{tgt}}\in\mathbb{R}^{L\times d}$, $L$ is the latent length, and $d=128$ is the latent dimension. The instruction $c$ is encoded by a pretrained language encoder and injected into the editor through cross-attention.

We use a Diffusion Transformer backbone \citep{peebles2023scalable} because spatial editing requires coordination across events, temporal regions, and FOA channels. Since $z_{\mathrm{src}}$ and the noisy target latent have the same temporal length, we concatenate them along the feature dimension before the Transformer input projection. The self-attention layers can then learn frame-level correspondences between the source scene and the latent being generated, while cross-attention supplies the instruction condition. The editor is trained with conditional flow matching \citep{lipman2023flow,guo2025techsinger}, which matches the one-stage source-to-target supervision by learning a vector field from noise to the target edited latent. Let $z_1=z_{\mathrm{tgt}}$ be the target latent and $z_0\sim\mathcal{N}(0,I)$ be a Gaussian noise latent. For a sampled timestep $t\sim\mathcal{U}(0,1)$, we define
\begin{equation}
    z_t = (1-t)z_0 + t z_1, \quad v=z_1-z_0.
\end{equation}
We optimize the following flow-matching objective to regress the target velocity:
\begin{equation}
    \mathcal{L}_{\mathrm{FM}}
    =
    \mathbb{E}_{z_0,t}
    \left[
    \left\|
    v - v_{\theta}(z_t,t,z_{\mathrm{src}},c)
    \right\|_2^2
    \right].
\end{equation}
The source latent gives the vector field an explicit preservation condition, while the instruction specifies the intended edit.

\subsection{Spatial Edit MoE with dual-level routing}

A single shared feed-forward module must cover very different editing behaviors: inserting an event, relocating a source, changing a trajectory, or modifying room acoustics. This is especially brittle for compound instructions, where several edit types interact. We introduce Spatial Edit Mixture-of-Experts (SE-MoE), a MoE module \citep{shazeer2017outrageously} in the flow-matching editor, to separate task-level routing from local edit routing. As illustrated in Figure~\ref{fig:wide}, SE-MoE replaces the feed-forward block in every Transformer layer.

SE-MoE contains three types of experts. Task experts are selected from the instruction representation and remain fixed for all latent frames in the same sample, giving task-aware expert combinations for different edit families and compound instructions. Routed experts are selected independently for each latent token, allowing local adaptation to edited regions and sound events. Null experts return zero tensors, allowing unchanged regions to avoid unnecessary transformations.

Given a latent hidden state $x_i$ at frame token $i$, the frame-level router computes $p_i=\mathrm{softmax}(W_g(x_i+W_t e_t))$, where $e_t$ is the flow-matching timestep embedding. Rather than using a fixed top-$k$ gate, we use top-$p$ routing and select the smallest expert set $\mathcal{S}_i$ whose cumulative probability exceeds a threshold $\rho$. Simple or unchanged frames can therefore activate fewer experts, while complex edited regions can use more experts.

For task-level routing, the instruction states are pooled with learned task queries to obtain instruction-conditioned expert probabilities $q(c)$. We apply the same top-$p$ selection to obtain the task expert set $\mathcal{T}(c)$ and normalize the selected probabilities as $\beta_e$. The final SE-MoE output combines frame-level routed experts and instruction-level task experts:
\begin{equation}
    \mathrm{SE\text{-}MoE}(x_i)
    \!=\!\!\!\!
    \sum_{e\in\mathcal{S}_i \cap \mathcal{E}_{r}}\!
    \alpha_{i,e} f^{r}_{e}(x_i)
    +\!
    \sum_{e\in\mathcal{T}(c)}
    \beta_{e} f^{t}_{e}(x_i)
\end{equation}
where $\mathcal{E}_{r}$ is the routed expert set, $f^{r}_{e}$ and $f^{t}_{e}$ denote routed and task experts, respectively, and $\alpha_{i,e}$ is the normalized frame-level routing weight. Null experts may be selected by the frame-level router, but they contribute zero output and are omitted from the summation.

We use a lightweight routing regularizer, denoted as $\mathcal{L}_{\mathrm{moe}}$, to prevent degenerate expert usage. Details are provided in the Appendix~\ref{append:se-moe}.

\subsection{Training strategy}

Training directly on complex editing triplets asks the model to learn spatial audio generation, language grounding, and edit execution at the same time. A failure on a compound instruction may therefore come from poor scene understanding rather than from the edit operation itself. We use a staged curriculum to reduce this ambiguity. First, text-to-FOA pretraining uses the scene-level FOA-caption pairs described above, giving the model basic spatial audio generation ability and grounding for audio events, spatial information, dynamic changes, and environmental information.

After pretraining, we initialize the editing model from the text-to-FOA model and fine-tune it on editing triplets $(a_{\mathrm{src}}, c, a_{\mathrm{tgt}})$. We first train on single-operation edits from the four editing axes so that the model learns atomic editing abilities. We then add multi-operation editing data for joint training. A compound edit may change several factors at the same time, such as replacing a source while moving it farther away or changing the room size while preserving all sound events. We supervise the model with the final target scene rather than a sequence of intermediate edits. This one-stage formulation avoids error accumulation and encourages the model to learn the joint acoustic effect of compound spatial edits.

\subsection{Spatial Preference Optimization}

Supervised flow matching specifies the target, but it does not explicitly contrast the target with plausible editing failures. In practice, the model can still copy the source audio, modify the wrong event, place an event in the wrong direction, or produce an implausible spatial transformation. We address these failures with Spatial Preference Optimization (SPO), a Direct Preference Optimization (DPO)-based objective \citep{rafailov2023direct, majumder2024tango2, wang2025t2a}, built from edit-specific negative targets.

For each edit type, we design negative samples that reflect typical model errors. For example, for extraction, the negative sample is generated by either extracting the wrong event or performing no operation. These two cases respectively discourage event misunderstanding and simple source copying. Similar negative designs are used for other edit types and are detailed in the appendix. In SPO data construction, wrong-edit negatives and no-operation negatives are generated at a ratio of $2:1$. For each edit type, we construct 10K SPO preference examples, with an average duration of approximately 9.17 seconds.

Let $z^{+}$ and $z^{-}$ denote the positive and negative target latents. For the policy model $\theta$, we compute the per-sample flow-matching losses
\begin{equation}
    \ell_{\theta}^{+}=\ell_{\theta}(z^{+},z_{\mathrm{src}},c), \quad
    \ell_{\theta}^{-}=\ell_{\theta}(z^{-},z_{\mathrm{src}},c).
\end{equation}
A frozen reference model $\theta_{\mathrm{ref}}$ gives the corresponding losses
$\ell_{\mathrm{ref}}^{+}$ and $\ell_{\mathrm{ref}}^{-}$. Since lower flow loss indicates a higher preference, we define
\begin{equation}
    \Delta =
    \beta
    \left[
    (\ell_{\theta}^{-}-\ell_{\theta}^{+})
    -
    (\ell_{\mathrm{ref}}^{-}-\ell_{\mathrm{ref}}^{+})
    \right],
\end{equation}
and optimize $\mathcal{L}_{\mathrm{SPO}}=-\log\sigma(\Delta)$.

We keep the positive supervised loss:
\begin{equation}
    \mathcal{L}_{\mathrm{align}}
    =
    \mathcal{L}_{\mathrm{SPO}}
    +
    \lambda_{\mathrm{pos}}\ell_{\theta}^{+}
    +
    \lambda_{\mathrm{moe}}\mathcal{L}_{\mathrm{moe}}.
\end{equation}
During SPO training, the positive and negative samples share the same noise and timestep, so the preference comparison focuses on the target edit rather than sampling randomness.

\subsection{Inference strategy}

At inference time, the source audio and instruction need different guidance strengths. The source condition should preserve unrelated content and spatial context, while the instruction condition should enforce the requested change. A single CFG scale on the joint condition cannot control these two effects separately. We therefore use a staircase classifier-free guidance (CFG) strategy \citep{ho2022classifier}. With condition dropout, each solver step can evaluate three velocity predictions:
\begin{equation}
\begin{aligned}
    v_{\varnothing} &= v_{\theta}(z_t,t,\varnothing,\varnothing), \\
    v_{\mathrm{src}} &= v_{\theta}(z_t,t,z_{\mathrm{src}},\varnothing), \\
    v_{\mathrm{all}} &= v_{\theta}(z_t,t,z_{\mathrm{src}},c).
\end{aligned}
\end{equation}
Here $v_{\varnothing}$ is the unconditional prediction, $v_{\mathrm{src}}$ uses only the input audio, and $v_{\mathrm{all}}$ uses both the input audio and instruction. The guided velocity is
\begin{equation}
    \tilde{v}_{\theta}
    =
    v_{\varnothing}
    +
    s_{\mathrm{src}}(v_{\mathrm{src}}-v_{\varnothing})
    +
    s_{\mathrm{inst}}(v_{\mathrm{all}}-v_{\mathrm{src}}),
\end{equation}
where $s_{\mathrm{src}}$ controls source preservation and $s_{\mathrm{inst}}$ controls edit strength. The staircase path $\varnothing \rightarrow z_{\mathrm{src}} \rightarrow (z_{\mathrm{src}},c)$ first anchors generation to the source scene and then applies the instruction as a residual edit. Compared with a flat two-scale CFG formulation used in multi-condition editing \citep{brooks2023instructpix2pix}, this separation makes the guidance easier to tune and reduces conflicts between preservation and instruction following. After integrating the guided flow from $t=0$ to $t=1$, the final edited FOA waveform is decoded as $\hat{a}_{\mathrm{tgt}}=\mathcal{D}(\hat{z}_{\mathrm{tgt}})$.

\section{Experiments}

\subsection{Experimental Setup}

\paragraph{Metrics}
We evaluate audio editing quality using both semantic and spatial metrics. Following prior audio generation and editing works~\citep{lan2025smartdj,wang2023audit}, we report FD, FAD, KL, and LSD. FD and FAD measure distributional distance in pretrained audio embedding spaces, KL measures event-level class-probability mismatch, and LSD measures spectral distortion between generated and reference audio. We further report CLAP score~\citep{elizalde2023clap} to evaluate instruction following, computed as the cosine similarity between the editing instruction and the generated audio in the CLAP embedding space.

For spatial fidelity, we draw on prior evaluations of localization and listener preference \citep{pan2025spatialeval} and evaluate all methods in a common two-channel stereo space for a fair comparison with stereo spatial audio editing baselines. Since our model directly generates FOA audio, we render both our FOA outputs and references into stereo audio using the same fixed FOA-to-stereo decoder before metric computation. Baseline methods are evaluated using their stereo outputs under the same stereo metric pipeline.
Following the stereo spatial metrics used in SmartDJ~\citep{lan2025smartdj}, we calculate GCC MSE (GCC) based on Generalized Cross-Correlation with Phase Transform (GCC-PHAT), and use StereoCRW~\citep{chen2022sound} to produce stereo audio features to evaluate CRW MSE (CRW) and Fréchet Stereo Audio Distance (FSAD).


\paragraph{Baselines}
Since no prior method directly edits FOA audio, we decode each FOA result into a stereo proxy when comparing with stereo baselines. 
The FOA signals already contain the PyRoom-simulated room response, and no additional RIR filtering is applied.
We compare with AudioEditor~\citep{jia2024audioeditor}, ZETA~\citep{manor2024zeta}, SDEdit~\citep{meng2022sdedit}, and SmartDJ~\citep{lan2025smartdj}. 
For AudioEditor, we replace its original Auffusion backend~\citep{xue2024auffusion} with the BEWO/SpatialSonic spatial-audio generator~\citep{sun2025bewo} to support binaural outputs. 
For SDEdit and ZETA, we follow SmartDJ and use Stable Audio Open~\citep{evans2025stable} as the generative backbone. 
For SmartDJ, we adapt it to a one-stage setting by feeding all editing operations to the editor jointly in a single inference pass.

\paragraph{Implementation Details}
We first convert the audio editing data into FOA-caption pairs to train a base text-to-FOA model. The base model is trained for 300K steps on 975K samples, enabling it to generate description-following FOA. We then initialize the editing model from the base model and train it on editing data for 500K steps, followed by 2 epochs of SPO post-training. During training, the source audio and text condition are independently replaced with empty inputs with a probability of 10\%, enabling the unconditional, source-only, and fully conditioned branches used by staircase CFG. We set $s_{src}$ and $s_{inst}$ as 3.0 in all experiments.
Unless otherwise specified, all model results are obtained from a single training run, and the reported objective metrics are averaged over the corresponding evaluation samples.

\subsection{Main results}

\begin{table*}[t]
\centering
\caption{Objective evaluation results averaged over edit types. HP denotes the pairwise human preference split between each baseline and our method in terms of audio quality and instruction-audio alignment.}
\label{tab:objective_metric}
\resizebox{\textwidth}{!}{
\begin{tabular}{lcccccccccc}
\toprule
Method & Speed & FD$\downarrow$ & FAD$\downarrow$ & KL$\downarrow$ & LSD$\downarrow$ & CLAP$\uparrow$ & GCC$\downarrow$ & CRW$\downarrow$ & FSAD$\downarrow$ & HP \\
\midrule
ZETA &  56.26s& 14.84&  4.03&  4.43&  3.29&  0.17&  51.85&  22.12&  0.78&  17.88\%-82.12\%\\
AudioEditor &  119.97s& 8.66&  3.19&  2.46&  2.89&  0.19&  28.19&  44.50&  0.55&  25.67\%-74.33\%\\
SDEdit &  32.28s&  8.60&  2.03&  2.41&  2.72&  0.18&  24.19&  25.48&  0.40&  23.30\%-76.70\%\\
SmartDJ &  4.23s& 8.47&  2.87&  2.69&  2.56&  0.17&  20.32&  33.07&  0.32&  38.44\%-61.56\%\\
Ours &  \textbf{1.17s}& \textbf{5.47}&  \textbf{1.16}&  \textbf{1.67}&  \textbf{1.07}&  \textbf{0.21}&  \textbf{12.39}&  \textbf{18.80}&  \textbf{0.31}&  -\\
\bottomrule
\end{tabular}
\vspace{-0.5em}
}
\end{table*}

\begin{table}[t]
\centering
\caption{Objective comparison on motion-related editing operations. Dis Motion means Distance Motion, indicating the distance of a sound event gradually changes.}
\label{tab:motion_ops}
\resizebox{\columnwidth}{!}{
\begin{tabular}{llccccccc}
\toprule
Operation & Method & FD$\downarrow$ & FAD$\downarrow$ & KL$\downarrow$ & LSD$\downarrow$ & GCC$\downarrow$ & CRW$\downarrow$ & FSAD$\downarrow$ \\
\midrule
\multirow{2}{*}{Angle Motion}
  & SmartDJ &  6.70&  2.40&  1.92&  2.36&  23.25&  50.16&  0.29\\
  & Ours    &  3.41&  1.23&  0.66&  0.80&  1.15&  14.84&  0.23\\
\midrule
\multirow{2}{*}{Dis Motion}
  & SmartDJ &  11.10&  1.88&  3.18&  2.77&  16.45&  28.66&  0.33\\
  & Ours    &  5.54&  0.93&  1.25&  0.86&  1.74&  11.32&  0.22\\
\bottomrule
\end{tabular}
}
\vspace{-0.6em}
\end{table}

\paragraph{Objective evaluation}

Table~\ref{tab:objective_metric} reports the average objective results over all edit types and multi-task editing. Our method achieves the best performance across efficiency, semantic quality, reconstruction quality, and spatial fidelity. In terms of inference speed, our model edits a sample in 1.17s on average, faster than all baselines. This is mainly because \textsc{SwanWeave} performs one-stage latent editing, while several baselines require many iterative diffusion sampling or sequential editing operations. Compared with the strongest baseline results, our model is better in FD, FAD, KL and LSD. These gains indicate higher-quality edited audio, better event consistency, and lower spectral distortion. Our method also obtains the highest CLAP score, suggesting stronger alignment with editing instructions.

For spatial fidelity, our method consistently outperforms all baselines on GCC, CRW, and FSAD. In particular, the lower GCC error shows that the directional structure encoded in FOA channels is better preserved after editing. The improvement on CRW further indicates more accurate stereo spatial cues after FOA-to-stereo decoding, while the best FSAD score suggests that our method produces spatial representations closer to the target distribution. These results show the advantage of directly modeling FOA audio instead of relying only on stereo-domain editing.

Table~\ref{tab:motion_ops} further evaluates motion-related editing operations. Compared with SmartDJ, our method achieves substantial gains on both angle motion and distance motion. For angle motion, GCC is reduced from 23.25 to 1.15 and CRW from 50.16 to 14.84. For distance motion, GCC is reduced from 16.45 to 1.74 and CRW from 28.66 to 11.32. These results show that our model is especially effective for dynamic spatial edits, where accurate temporal changes in direction and distance are required.

\paragraph{Human evaluations}
We further conduct pairwise human evaluations with three annotators. For each test case, annotators are given the source audio, the editing instruction, and two edited results, and are asked to choose the better one in terms of audio quality and instruction-audio alignment. The HP column in Table~\ref{tab:objective_metric} reports the preference ratio between each baseline and our method. Our method is preferred over ZETA, AudioEditor, SDEdit, and SmartDJ. The human evaluation results are consistent with the objective metrics and further confirm the perceptual quality and instruction-following ability of our method.

\subsection{Ablations}

\begin{table}[t]
\centering
\caption{Ablation study of different model components.}
\label{tab:ablation}
\resizebox{\columnwidth}{!}{
\begin{tabular}{lcccccccc}
\toprule
Variation 
& FD$\downarrow$ 
& FAD$\downarrow$ 
& KL$\downarrow$ 
& LSD$\downarrow$ 
& CLAP$\uparrow$ 
& GCC$\downarrow$ 
& CRW$\downarrow$ 
& FSAD$\downarrow$ \\
\midrule
w/o pretrain 
& 8.49 & 3.15 & 2.78 & 2.33 & 0.16 & 21.24 & 25.09 & 0.39 \\
\midrule
w/o MoE 
& 6.84 & 2.72 & 2.34 & 1.66 & 0.21 & 14.05 & 19.51 & 0.29 \\
w/o task expert 
& 6.21 & 2.14 & 1.98 & 1.24 & 0.18 & 13.76 & 17.91 & 0.38 \\
\midrule
w/o SPO 
& 5.94 & 1.82 & 1.85 & 1.10 & 0.21 & 15.45 & 19.89 & 0.32 \\
\bottomrule
\end{tabular}
}
\end{table}

\begin{table*}[t]
\centering
\caption{Comparison between one-stage and multi-stage editing on compound-edit samples.}
\label{tab:one_multi}
\resizebox{\textwidth}{!}{
\begin{tabular}{lccccccccc}
\toprule
Method
& Speed (s)$\downarrow$
& FD$\downarrow$
& FAD$\downarrow$
& KL$\downarrow$
& LSD$\downarrow$
& CLAP$\uparrow$
& GCC$\downarrow$
& CRW$\downarrow$
& FSAD$\downarrow$ \\
\midrule
SmartDJ multi-stage
& 10.73 & 9.88 & 3.12 & 2.82 & 2.99 & 0.17 & 27.99 & 47.99 & 0.44 \\
SmartDJ one-stage
& 4.23 & 8.47 & 2.87 & 2.69 & 2.56 & 0.17 & 20.32 & 33.07 & 0.32 \\
\midrule
Ours multi-stage
& 3.96 & 7.16 & 1.67 & \textbf{1.36}& 1.82 & \textbf{0.23} & 19.45 & 22.29 & 0.37 \\
Ours one-stage
& \textbf{1.17} & \textbf{5.47} & \textbf{1.16} & 1.67 & \textbf{1.07} & 0.21
& \textbf{12.39} & \textbf{18.80} & \textbf{0.31} \\
\bottomrule
\end{tabular}
}
\end{table*}
\paragraph{Ablation of pretraining}
As shown in Table~\ref{tab:ablation}, removing FOA-caption pretraining and training directly on the editing dataset leads to clear degradation across all metrics. FD increases from 5.47 to 8.49, KL increases from 1.67 to 2.78, and CLAP decreases from 0.21 to 0.16. This indicates that text-to-FOA pretraining provides a strong initialization for FOA generation and improves semantic alignment, reconstruction quality and spatial accuracy before editing-specific training.

\paragraph{Ablation of SE-MoE with dual-level routing}
We further analyze the effectiveness of the proposed SE-MoE module. Removing the entire MoE block leads to degradation on FD, KL, and LSD, showing that a single shared feed-forward module is less effective for modeling heterogeneous spatial editing operations. This confirms the benefit of expert specialization for different edit patterns and temporal regions.
We then remove only the task experts while keeping the frame-level routed and null experts. This variant performs better than the model without MoE, indicating that token-level routing is useful for local temporal adaptation. However, it is worse than the full model. This suggests that caption-conditioned task experts provide complementary instruction-level specialization. Overall, the ablation verifies that both frame-level routed experts and task-level experts are important for effective FOA spatial audio editing.

\paragraph{Ablation of SPO post-training}
Finally, we evaluate the effect of the proposed SPO post-training stage by removing it from the full training pipeline, and it is consistently worse than the full model in FD, FAD, KL, and LSD. The spatial metrics show a similar trend: removing SPO increases GCC and CRW, while FSAD also increases slightly. This indicates that supervised training alone does not fully suppress plausible editing failures, such as source copying, wrong-event modification, or incorrect spatial placement. After SPO post-training, the model better distinguishes the desired target edit from edit-specific negative targets. The improvement is especially reflected in distributional and event-level metrics. These results show that SPO provides complementary preference-level supervision beyond the positive flow-matching objective and improves perceptual editing quality and instruction following.

\paragraph{One-stage vs. Multi-stage Editing}
To directly examine the effect of one-stage modeling for compound editing, we compare one-stage and multi-stage inference on compound-edit samples. We select 100 test examples each with two, three, and four editing operations. In the multi-stage setting, the compound edit is decomposed into multiple operations and executed sequentially, whereas the one-stage setting directly predicts the final edited result from the compound instruction.

As shown in Table~\ref{tab:one_multi}, sequential execution generally degrades audio and spatial fidelity while substantially increasing inference time. For our model, one-stage editing reduces FD from 7.16 to 5.47, FAD from 1.67 to 1.16, and LSD from 1.82 to 1.07, while also improving GCC, CRW, and FSAD from 19.45, 22.29, and 0.37 to 12.39, 18.80, and 0.31, respectively. Similar degradation is observed for SmartDJ under multi-stage execution. These results suggest that repeatedly editing intermediate outputs can accumulate reconstruction and spatial errors, whereas directly predicting the final target scene better preserves the overall acoustic and spatial structure of compound edits. Meanwhile, the mixed trends on KL and CLAP indicate that the primary advantage of one-stage modeling is improved overall reconstruction and spatial consistency rather than uniformly improving every semantic metric.

\section{Conclusion}

We present \textsc{SwanWeave}, a one-stage framework for instruction-guided 3D FOA spatial audio editing. 
Using paired FOA editing supervision built with controllable room simulation, \textsc{SwanWeave} supports diverse edits over audio events, spatial cues, temporal dynamics, and environmental properties. 
The model combines latent flow-matching editing, SE-MoE, and SPO to handle heterogeneous instructions and improve preference alignment. 
Experiments show consistent improvements over general audio editors and spatial audio baselines across objective and human preference metrics. 
These results highlight the promise of directly modeling FOA spatial audio for controllable 3D audio creation.

\newpage

\section*{Limitations}
Although \textsc{SwanWeave} shows strong performance on instruction-guided FOA spatial audio editing, several limitations remain. First, our paired supervision is constructed with controllable room simulation. This provides accurate source-target alignment for training, but may not fully capture the acoustic complexity of real recordings, such as microphone characteristics, background noise, occlusion, and highly irregular room responses. Extending the framework to real recorded FOA data is an important direction for future work.

Second, while our dataset covers diverse editing operations over audio events, spatial information, dynamic changes, and environmental information, the current scenes are still limited in duration and complexity. Most examples contain a small number of sound events and are approximately 10 seconds long. Editing longer scenes with denser event mixtures, overlapping trajectories, and more open-ended instructions may require stronger long-context modeling and more diverse supervision.

\section*{Ethical considerations}
Our training data is constructed from open-source speech and sound-effect corpora, together with controllable room simulation. We use these data for modeling spatial editing operations rather than for imitating specific speaker identities. Nevertheless, speech data may contain human voices, and future extensions to real recorded FOA data should carefully consider consent, privacy, and dataset licensing. The simulated-data construction also reduces the need to collect private real-world spatial recordings.

\section*{Acknowledgments}

This work was supported by the National Natural Science Foundation of China under Grant No.~U25B2064.

\bibliography{latex/custom,latex/spatial_edit_refs}

\newpage
\appendix

\section{Dataset Details}
\subsection{Detailed implementation of spatial audio}
We construct an editing training dataset from open-source speech and sound-effect corpora, including AudioCaps \citep{kim2019audiocaps}, FSDKaggle2019 \citep{fonseca2019audio}, PicoAudio \citep{xie2024picoaudio}, LibriSpeech \citep{panayotov2015librispeech}, and Spatial LibriSpeech \citep{sarabia2023spatial}. We follow common audio-data practices for source collection, transcription, annotation, and coverage analysis \citep{zhang2024gtsinger,guo2025stars,li2024robust,pan2025syntheticsingers}. For each example, we sample two to four audio events to form a source scene and synthesize its target scene according to the requested edit. This source-target construction also follows the use of paired supervision for controllable audio generation under explicit content and style conditions \citep{zhang2025versatile,zhang2025tcsinger,zhang2024tcsinger,zhang2024stylesinger}.


We employ PyRoomAcoustics~\citep{scheibler2018pyroomacoustics} for controllable room simulation of FOA spatial audio. Specifically, the simulation allows us to explicitly control the source direction, source-listener distance, motion trajectory, and room reflection conditions. We do not adopt Habitat\citep{savva2019habitat} for rendering, as its complex scene geometry may cause occlusions in direct sound propagation and introduce strong multipath reflections, leading to inaccurate directional cues at the listener position. In contrast, PyRoomAcoustics provides a more direct and controllable simulation framework, and is therefore used for generating our spatial audio data. The default room size in our dataset is $[21, 21, 10]$ meters, corresponding to length, width, and height, respectively.

We render FOA spatial audio using PyRoomAcoustics. For each event, we construct a shoebox room and synthesize a room impulse response from the source position to a microphone placed at the listener(the center of the room). The RIR captures the direct path, early reflections, and late reverberation determined by room geometry, wall absorption, and reflection order. We then encode the center RIR into first-order Ambisonics using analytic directional gains derived from the source-listener unit vector. The dry event signal is convolved with the resulting four-channel FOA RIR and mixed into the scene timeline. Unlike binaural rendering, which synthesizes separate left and right ear responses, our rendering produces an Ambisonic sound field in the W,Y,Z,X channel order, allowing the model to learn editable spatial cues in a FOA representation.

\paragraph{Direction description}
Directional information is one of the most important distinctions between FOA and conventional mono audio. In our dataset, we define ten spatial directions: right, front, left, back, front-right, front-left, back-left, back-right, above, and below. The first eight correspond to azimuth angles of $[0^\circ, 90^\circ, 180^\circ, 270^\circ, 45^\circ, 135^\circ, 225^\circ, 315^\circ]$, while the above and below directions indicate that the source is located vertically relative to the listener, without being restricted to directly overhead or directly beneath. To introduce variability, after specifying the azimuth and distance, we randomly sample a point within a small spherical region centered at this nominal position as the final source location, while keeping the primary directional category fixed.

\subsection{Basic edit actions}
In this subsection, we explain the meaning of each single-step edit action, as well as how the corresponding training targets and negative samples for preference learning are constructed.

\paragraph{Add}
Add a new sound event to the original audio. 
The corresponding negative samples either add an incorrect event or perform no operation by copying the original audio.

\paragraph{Remove}
Remove a specified sound event from the original audio while preserving the remaining sound events and spatial information. 
The corresponding negative samples either remove an incorrect event or perform no operation by copying the original audio.

\paragraph{Replace}
Replace a specified sound event in the original audio with a new sound event.
The corresponding negative samples either replace an incorrect event or perform no operation by copying the original audio.

\paragraph{Extract}
Extract the specified sound event from the original audio and remove the other sound events.
The corresponding negative samples either extract an incorrect event or perform no operation by copying the original audio.

\paragraph{Volume increase/Volume fade}
Increase or decrease the volume of a specified sound event in the original audio while keeping the other sound events unchanged as much as possible.
The corresponding negative samples either operate on an incorrect event or perform no operation.

\paragraph{Angle change}
Change the angular position of a specified sound event in the original audio.
The corresponding negative samples either move the event to an incorrect angle, apply the angle change to an incorrect event, or perform no operation.

\paragraph{Distance change}
Change the distance of a specified sound event in the original audio.
The corresponding negative samples either apply the distance change to an incorrect event or perform no operation.

\paragraph{Room change}
Change the room size or reverberation condition of the original audio while keeping the sound event content and spatial layout unchanged as much as possible.
We define four room-size categories by length, width, and height: small room [2, 2, 2], medium room [6, 6, 4], large room [21, 21, 10], and extra-large room [100, 100, 100], where the extra-large room is regarded as an open-space scene.
The corresponding negative samples either perform no room change or change the scene to an incorrect room condition.

\paragraph{Angle motion}
Move a specified sound event dynamically from an initial direction to a target direction in the original audio.
The corresponding negative samples either place the event directly at the target position without motion, move it toward an incorrect direction, or perform no operation.

\paragraph{Distance motion}
Move a specified sound event dynamically from an initial distance to a target distance in the original audio.
The corresponding negative samples either keep the event static, move it to an incorrect position or with an incorrect distance trajectory, or perform no operation.

\begin{figure}[t]
    \centering
    \includegraphics[width=\linewidth]{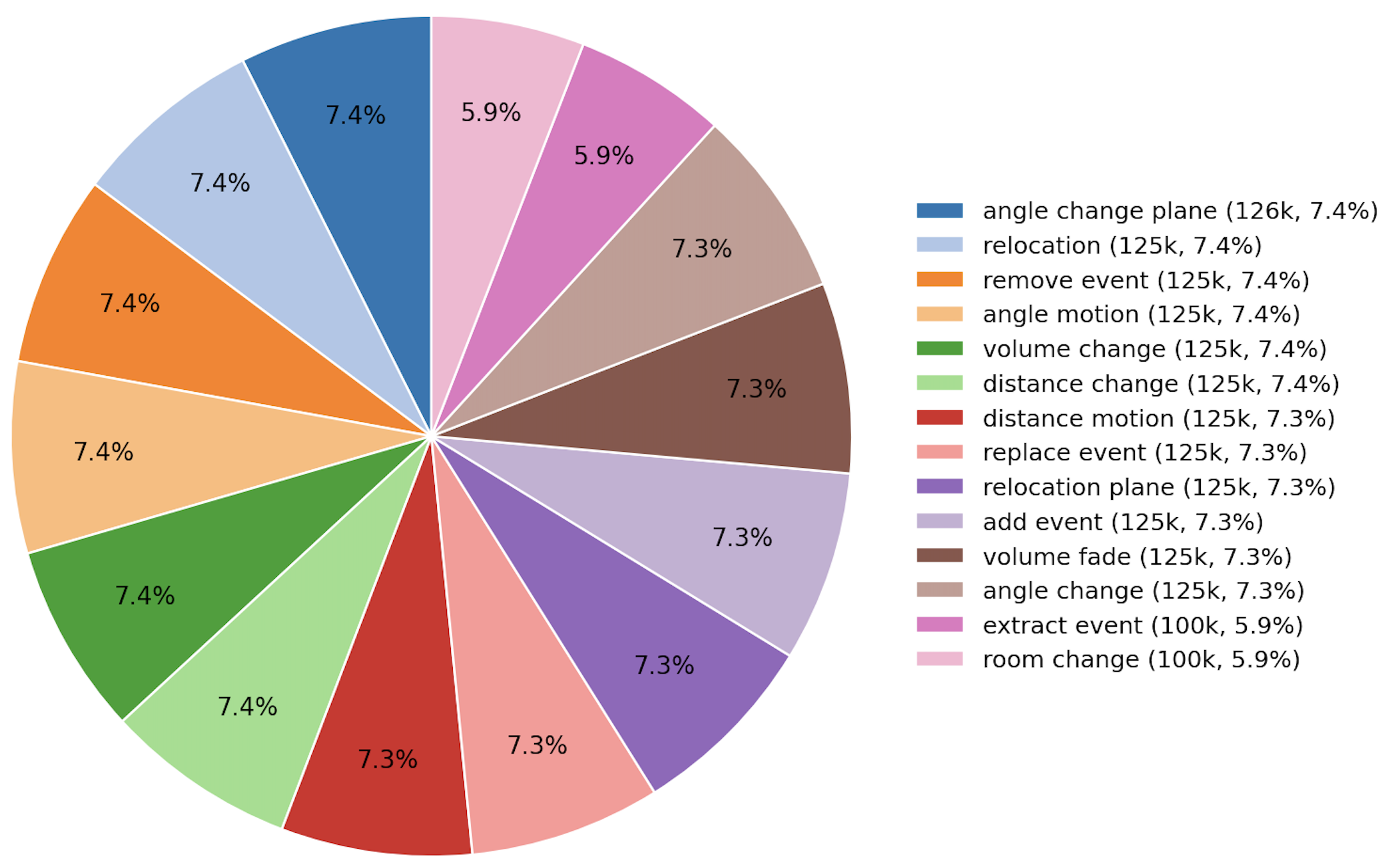}
    \caption{Overview of the training data distribution. Most edit types contain about 125K examples.}
    \label{fig:data distribution}
\end{figure}

The data distribution of different single-operation edit actions is shown in Figure~\ref{fig:data distribution}.

\subsection{Prompt for edit instructions}
We use Gemini 2.5 Pro to generate the instructions required for editing. The prompt used is as follows:

\begin{tcolorbox}[
    colback=gray!3,
    colframe=black!50,
    boxrule=0.5pt,
    arc=2pt,
    left=6pt,
    right=6pt,
    top=6pt,
    bottom=6pt,
    title={Prompt for Editing Instruction Generation}
]
\small\ttfamily
You are a precise spatial audio editor. Given a specific edit operation, write a short, natural English spoken instruction.

STATIC edits: ADD, REMOVE, REPLACE, VOLUME CHANGE, DISTANCE CHANGE, ANGLE CHANGE, RELOCATION.

DYNAMIC edits: DISTANCE MOTION, ANGLE MOTION, VOLUME FADE.

FILTER edit: EXTRACT.

ENVIRONMENT edit: ROOM CHANGE.

For DYNAMIC edits, the caption MUST convey continuous motion/change, not a static state.

For STATIC edits, describe the resulting state, not motion.

For ROOM CHANGE, describe the change of room size.

Each caption must be 10--25 words, one sentence. Output exactly 3 lines, one caption per line, no numbering, no quotes.

\vspace{0.5em}
Room: \{room\}

Scene before editing:

\{original\}

Edit: \{edit\_detail\}

Write 3 different English instructions for this exact edit.
\end{tcolorbox}

\section{Model Details}

\subsection{SE-MoE routing regularization}
\label{append:se-moe}

This section gives the routing regularizer used for SE-MoE. The main text keeps only the compact notation $\mathcal{L}_{\mathrm{moe}}$ because these terms stabilize expert usage but are not the central modeling contribution.

For the frame-level router, let $u^{r}_{b,i,e}$ be the routing logit for expert $e$ at latent token $i$ of sample $b$, and let
\begin{equation}
    p^{r}_{b,i,e}
    =
    \frac{\exp(u^{r}_{b,i,e})}{\sum_{e'}\exp(u^{r}_{b,i,e'})}
\end{equation}
be the corresponding routing probability. The selected top-$p$ expert set is denoted by $\mathcal{S}_{b,i}$. For the task-level router, let $u^{t}_{b,e}$ and $q_{b,e}$ denote the task-router logit and probability for expert $e$, and let $\mathcal{T}_{b}$ be the selected task expert set.

We use a load-balancing loss to avoid routing collapse. For the frame-level router, expert importance and expert load are
\begin{equation}
    I^{r}_{e}=\sum_{b,i} p^{r}_{b,i,e},
    \qquad
    L^{r}_{e}=\sum_{b,i} \mathbb{I}[e\in\mathcal{S}_{b,i}].
\end{equation}
For the task-level router, they are
\begin{equation}
    I^{t}_{e}=\sum_{b} q_{b,e},
    \qquad
    L^{t}_{e}=\sum_{b} \mathbb{I}[e\in\mathcal{T}_{b}].
\end{equation}
Given a vector $x$, define
\begin{equation}
    \mathrm{CV}^{2}(x)=\frac{\mathrm{Var}(x)}{\mathrm{Mean}(x)^2+\epsilon}.
\end{equation}
The balancing term is
\begin{equation}
    \mathcal{L}_{\mathrm{bal}}
    =
    \mathrm{CV}^{2}(I^{r})
    +
    \mathrm{CV}^{2}(L^{r})
    +
    \mathrm{CV}^{2}(I^{t})
    +
    \mathrm{CV}^{2}(L^{t}).
\end{equation}

We also use a router z-loss to keep routing logits numerically stable:
\begin{equation}
\begin{aligned}
    \mathcal{L}_{z}
    =&\frac{1}{|\mathcal{B}|L}
    \sum_{b,i}
    \left(\log\sum_{e}\exp(u^{r}_{b,i,e})\right)^2 \\
    &+\frac{1}{|\mathcal{B}|}
    \sum_{b}
    \left(\log\sum_{e}\exp(u^{t}_{b,e})\right)^2,
\end{aligned}
\end{equation}
where $|\mathcal{B}|$ is the batch size and $L$ is the latent length.

For the null expert, we use the synthetic edit metadata to obtain a binary temporal mask $m_{b,i}$, where $m_{b,i}=1$ indicates that latent token $i$ overlaps an edited region and $m_{b,i}=0$ indicates an unchanged region. Let $e_{\varnothing}$ denote the null expert. The null-routing loss is
\begin{equation}
\begin{aligned}
    \mathcal{L}_{\mathrm{null}}
    =
    -\frac{1}{|\mathcal{B}|L}
    \sum_{b,i}
    \big[ &(1-m_{b,i})\log(p^{r}_{b,i,e_{\varnothing}}+\epsilon) \\
    &+m_{b,i}\log(1-p^{r}_{b,i,e_{\varnothing}}+\epsilon)
    \big].
\end{aligned}
\end{equation}
This term encourages unchanged regions to use the null expert while discouraging edited regions from routing to it.

The final regularizer used in training is
\begin{equation}
    \mathcal{L}_{\mathrm{moe}}
    =
    \lambda_{\mathrm{bal}}\mathcal{L}_{\mathrm{bal}}
    +
    \lambda_{z}\mathcal{L}_{z}
    +
    \lambda_{\mathrm{null}}\mathcal{L}_{\mathrm{null}}.
\end{equation}

\subsection{Negative targets for SPO}

SPO uses edit-specific negative targets rather than arbitrary corrupted audio. We construct two broad types of negatives. A \emph{wrong-edit} negative applies a plausible but incorrect edit, such as changing the wrong event, moving a source to the wrong direction, using the wrong motion trajectory, or applying the wrong room-acoustic condition. A \emph{no-operation} negative keeps the source scene unchanged. During SPO data construction, wrong-edit and no-operation negatives are mixed with a ratio of $2:1$, so the model is trained both to avoid semantically wrong edits and to avoid copying the source when an edit is required.

\section{More experiments}
\subsection{Selection of CFG scale}
\label{sec:cfg_scale}

We select the source and instruction CFG scales, $s_{\mathrm{src}}$ and $s_{\mathrm{inst}}$, by balancing source preservation and instruction adherence. Specifically, we evaluate 40 generated samples for each scale combination and manually assess whether the unedited source content is preserved and whether the requested edit is successfully performed. As shown in Table~\ref{tab:cfg_scale}, we set $s_{\mathrm{src}}=3.0$ and $s_{\mathrm{inst}}=3.0$ for all experiments, as this configuration achieves the best overall balance between preservation and editing success.

\begin{table}[t]
\centering
\caption{Selection of CFG scales. Each entry reports the number of samples with preserved unedited content / successful editing, out of 40 samples.}
\label{tab:cfg_scale}
\begin{tabular}{c|ccc}
\toprule
$s_{\mathrm{src}} \backslash s_{\mathrm{inst}}$
& 1.5 & 3.0 & 4.5 \\
\midrule
1.5 & 36 / 23 & 33 / 28 & 28 / 30 \\
3.0 & 36 / 24 & \textbf{36 / 31} & 30 / 31 \\
4.5 & 38 / 20 & 33 / 29 & 31 / 33 \\
\bottomrule
\end{tabular}
\end{table}

\subsection{Instruction robustness and complexity-stratified evaluation}
We further evaluate SPATIALEDIT on 50 instructions with colloquial, redundant, or ambiguous expressions, achieving 44\% instruction adherence and 76\% source preservation. We also stratify human evaluation by instruction complexity, where SPATIALEDIT achieves an average preference of 62\% over SmartDJ and shows a larger advantage as editing complexity increases.

\subsection{Human evaluation details}
For the human evaluation, annotators were presented with an original spatial audio clip, an editing instruction, and two edited versions, denoted as A and B. They were asked to choose the version they considered better by jointly considering the perceptual audio quality and how well the edited result followed the given instruction.

The exact instruction shown to the annotators was:

\begin{quote}
\textit{You will hear an original spatial audio clip, an editing instruction, and two edited versions (A and B). Please choose the one you find better, taking into account both the audio quality and how well the edit follows the instruction.}
\end{quote}
The annotators were volunteer graduate students recruited from our research group. They did not receive monetary compensation for participating in the evaluation.

\end{document}